\documentclass[aps,twocolumn,showpacs,groupedaddress]{revtex4}
\usepackage{graphicx}

\begin{document}

\title{Optimal Conditions for Atomic Homodyne Detection on Bose-Einstein Condensates}

\author{B. R. da Cunha and M. C. de Oliveira}

\affiliation{Instituto de F\'{\i}sica ``Gleb Wataghin'',
Universidade Estadual de Campinas, 13083-970, Campinas-SP, Brazil}

\date{\today}

\begin{abstract}

The dynamics of a two-mode Bose-Einstein condensate trapped in a
double-well potential results approximately in an effective Rabi
oscillation regime of exchange of population between both wells
for sufficiently strong overlap between the modes functions.
Facing this system as a temporal atomic beam splitter we show that
this regime is optimal for a nondestructive atom-number
measurement allowing an atomic homodyne detection, thus yielding
indirect relative phase information about one of the two-mode
condensates.

\end{abstract}

 \maketitle

\section{Introduction}
Since the first experimental achievements of Bose-Einstein
condensation with a diluted atomic gas \cite{bec1,bec2,bec3}, the
investigation related to the detection of condensate relative
phases and more specifically to the actual condensate quantum
state determination has generated a considerable literature
\cite{castin,leggett,walls}. A significant amount of discussion
has been directed towards the detection of the relative phase of
two Bose-Einstein Condensates (BECs), either in the form of
interference between two independent BECs
\cite{castin,andrews,hall,anderson} or due to indirect light probe
of independent condensates \cite{saba}. Essentially, acquiring
information about the BEC relative phase enables one to observe
many interesting dynamical phenomena such as Josephson effect
\cite{anderson,1,2,3,4,5,6,7,raghavan1,raghavan2,cataliotti,albiez}
and the transition from superfluid to Mott insulator \cite{bloch}.
However the knowledge of such a phase could also be employed for
partial or total inference on the condensate quantum state,
through tomographic reconstruction as proposed by many authors in
the past \cite{8,9,10}.

Indeed, an atomic interferometric device implemented with high
controllable parameters can be envisaged as an atomic
beam-splitter \cite{castin,adams,ijmpb}, which would be certainly
important for schemes of quantum state reconstruction of BECs
\cite{8,9,10}. It was previously noticed in Ref. \cite{marcos}
that a two-mode BEC trapped in a double-well potential could be
envisaged as the atomic version of the Balanced Optical Homodyne
Detection (BOHD), where the coherent tunneling of atoms would play
the role of a temporal atomic beam-splitter.  As it is well known,
BOHD consists of mixing the signal field with a coherent local
oscillator (LO) on a 50:50 beam-splitter to yield the necessary
phase-sensitivity for signal field quadratures detection
\cite{1a,milburnli}. In the same sense, the phase of a signal
atomic BEC mode could be thus determined by counting the
difference of atoms in the wells of the trapping potential
\cite{marcos}. However in Ref. \cite{marcos} only an approximate
description of a two-mode BEC was given, by neglecting
cross-collision between atoms in different wells. As such, the
approximate calculations were valid only for a small number of
atoms (small condensate) and short time, determined by the ratio
between tunneling and collision frequencies. Recently it was
demonstrated that cross-collisions between atoms trapped in a
double well potential could significantly increase the atom
tunneling rate for special trap configurations leading to an
effective linear Rabi regime of population oscillation between the
trap wells \cite{albiez,bruno}. This regime of oscillation is
optimal for atomic homodyne detection of a signal BEC phase if the
number of atoms in each well can be inferred from available
experimental techniques. The Josephson coupling between distinct
modes of a BEC trapped in a double well potential can be thus
regarded as a reliable system for realizing a temporal atomic
beam-splitter \cite{ijmpb} and consequently atomic homodyne
detection.

In this paper we develop the procedures for detecting a condensate
relative phase through atomic homodyne detection. For the proper
atomic homodyne detection process it is needed a secondary
detection process able to determine the difference of atoms in the
trap wells. The approach which we believe is the most promising in
measuring the phase of a condensate is based on an extension of
the homodyne measurements on a BEC proposed by Corney and Milburn
in Ref. \cite{4}. One of the wells of the double-well system is
placed inside an optical cavity, which is far off resonance with
respect to any dipole transition in the atomic sample, allowing a
dispersive interaction between the light field and the atomic gas.
Hence, the effect of the atoms is to shift the phase of the cavity
field by a given amount dependent on the balance of bosons in both
wells, which may be measured by homodyne interferometry reflecting
the internal dynamics of the condensate. Consequently, we simulate
the homodyne current and its relation to the quadrature phase,
possibly foreseeing experimental measured quantities and showing
how the presence of cross-collisions enables a dynamical regime
ideal for such homodyne interferometry scheme. This paper is
organized as follows. In Sec. II we review the main aspects
presented in Ref. \cite{bruno} for deriving the effective stable
Rabi regime. In Sec. III we derive a detection model based on
optical homodyne detection. In Sec. IV approximate solutions for
the equations discussed in Sec. III are given, allowing the
determination of the condensate relative phase through BOHD. In
SEC. V the back-action on the condensate phase due to the
continuous measurement process is analyzed and finally in Sec. VI
a conclusion encloses the paper.

\section{Effective Rabi regime and BEC quadratures detection}

Before proceeding, we present briefly the quantum dynamics of a
BEC trapped in a double well potential as derived in \cite{bruno}
in order to justify the homodyne measurements discussed in later
sections. The condensate model used in the following discussion
has been studied in previous papers \cite{6,7,bruno} and so we
only present an overview of it. Consider a double well potential
trapping a Bose-Einstein condensate. The potential barrier is
considered to be symmetric and the chemical potential is such that
only two single-particle states are bellow the barrier separating
the two wells, but in such a way that cross collisions between
bosons of both wells may not be negligible. Those hypotheses
enable a treatment of the many-body problem within a two-mode
approximation. The well known bosonic many body Hamiltonian in the
interaction picture is
\begin{eqnarray}
\hat{H}&=&\int d^{3}r\hat{\Psi}^{\dag}({\bf
r})\left[-\frac{\hbar^{2}}{2m}\nabla^{2}+V({\bf
r})\right]\hat{\Psi}({\bf r})\nonumber\\&&+\frac{U_{0}}{2}\int
d^{3}r\hat{\Psi}^{\dag}({\bf r})\hat{\Psi}^{\dag}({\bf
r})\hat{\Psi}({\bf r})\hat{\Psi}({\bf r})
\end{eqnarray}
where $m$ is the atomic mass, $U_{0}=\frac{4\pi\hbar^{2}a}{m}$
measures the strength of the two-body interaction, $a$ is the
s-wave scattering length, $\hat{\Psi}^{\dag}$ and $\hat{\Psi}$ are
the Heisenberg picture field operators. If we consider a dilute
gas in order that only s-scattering interactions are not
negligible  we may define the tunneling ($\Omega$), self-collision
($\kappa$), and cross-collision ($\eta$, $\Lambda$) rates,
respectively as
\begin{eqnarray}
\Omega&=&\frac{2}{\hbar}\int d^{3}{\bf r}u^{*}_{1}({\bf
r})[V(r)-\widetilde{V}^{(2)}({\bf r}-{\bf r}_{1})]u_{2}({\bf r}),\\
\kappa&=&\frac{U_{0}}{2\hbar}\int d^{3}{\bf r}|u_{i}|^{4},\\
\eta&=&(\frac{U_{0}}{2\hbar})\int d^{3}{\bf
r}u_{i}^{*}u_{i}u_{j}^{*}u_{j}, \\
\Lambda&=&(\frac{U_{0}}{2\hbar})\int d^{3}{\bf
r}u_{j}^{*}u_{j}u_{i}^{*}u_{i},
\end{eqnarray}
where $V^{(2)}$ is the harmonic approximation of the trapping
potential around each minimum and $u_{i}$ is the i-th mode
function such that $\hat{a}_i(t)=\int d^3{\bf r} u^*_i({\bf
r})\hat{\psi}({\bf r},t)$. It is then possible to write down a
two-mode single-particle Hamiltonian as
\begin{eqnarray}
\hat{H}&=&\hbar(2\Lambda(N-1)+\Omega)[\hat{a}^{\dag}\hat{b}+\hat{b}^{\dag}\hat{a}]+\hbar\eta[\hat{a}^{\dag}\hat{b}+\hat{b}^{\dag}\hat{a}]^{2}\nonumber\\
&&
+\hbar(\kappa-\eta)[(\hat{a}^{\dag})^{2}(\hat{a}^{2})+(\hat{b}^{\dag})^{2}(\hat{b}^{2})],
\end{eqnarray}
where we have used the bosonic field operators relation to
isomorphically map the many-body problem into the single-particle
one. If we now introduce the Schwinger angular momentum
representation
\begin{eqnarray}
\hat{J_{x}}&=&\frac{1}{2}(b^{\dag}b-a^{\dag}a),\\
\hat{J_{y}}&=&\frac{i}{2}(b^{\dag}a-a^{\dag}b),\\
\hat{J_{z}}&=&\frac{1}{2}(a^{\dag}b+b^{\dag}a),
\end{eqnarray} the Hamiltonian in Eq. (6) then becomes
\begin{equation}
\hat{H}=\hbar[\Omega+2\Lambda(N-1)]\hat{J}_{z}+4\hbar\eta\hat{J}_{z}^{2}+2\hbar(\kappa-\eta)\hat{J}_{x}^{2}
\end{equation}
where we have neglected terms proportional to N and $N^{2}$ since
they correspond only to a shift in the energy scale. The Casimir
invariant is
\begin{equation}
\hat{J}^{2}=\frac{\hat{N}}{2}(\frac{\hat{N}}{2}+1),
\end{equation}
which is analogous to an angular momentum model with total
eigenvalue given by $j=N/2$. We may now use the Heisenberg picture
to write the equations of motion for the angular momentum
operators as follows
\begin{eqnarray}
\dot{\hat{J}_{x}}&=&-\hbar[\Omega+2\Lambda(N-1)]\hat{J}_{y}-4\hbar\eta[\hat{J}_{y},\hat{J}_{z}]_{+},\\
\dot{\hat{J}_{y}}&=&\hbar[\Omega+2\Lambda(N-1)]\hat{J}_{x}-2\hbar(\kappa-3\eta)[\hat{J}_{z},\hat{J}_{x}]_{+},\\
\dot{\hat{J}_{z}}&=&2\hbar(\kappa-\eta)[\hat{J}_{y},\hat{J}_{x}]_{+},
\end{eqnarray}where $[\cdot,\cdot]_+$ are anticommutators.
This system of differential equations can be solved numerically,
and show a number of interesting effects, such as self-trapping as
discussed in Ref. \cite{bruno} or in Refs.
\cite{6,7,raghavan1,raghavan2} in absence of cross-collision terms
($\eta=\Lambda=0$). Eqs. (12-14) essentially show that when the
cross collision between the localized modes is taken into account
the mode volume is increased. For a given fixed number of
particles that means that the atomic density at each well is
decreased and so the
 self-collisions rate occurring in each mode, as given by $\kappa-\eta$. Also the
tunneling rate is increased as a consequence of the mode volume
increase \cite{bruno} as given by $\Omega+2\Lambda(N-1)$, and so
dependent not only on the cross-collisional rate $\Lambda$ but
also on the number of atoms in the trap.

It is easily seen from the system of equations (12-14) that the
presence of cross-collision inhibits self-trapping in the limit
where $\kappa-\eta\ll\Omega+2\Lambda(N-1)$, especially when
$\eta\rightarrow\kappa$. In such a case, Eqs. (12-14) result in
\begin{eqnarray}
 \dot{\hat{J}_{x}}&=&-\hbar\Omega'\hat{J}_{y},\\
\dot{\hat{J}_{y}}&=&\hbar\Omega'\hat{J}_{x},\\
\dot{\hat{J}_{z}}&\approx&0,
\end{eqnarray} being thus $\hat{J}_{z}$ approximately constant of motion.
Here we have defined
$\Omega'\equiv\Omega+2\Lambda(N-1)+8\kappa\hat{J}_{z}(0)$ as the
new tunneling frequency, which explicitly depends on the
cross-collision rate, on $N$, and on the initial condition for
$\hat{J}_{z}$. This regime could in principle be attained as
discussed in Ref. \cite{bruno} for special trap configurations.
The new set of equations is thus easily solved to give
\begin{equation}
\hat{J}_{x}(t)=\hat{J}_{x}(0)\cos\Omega't+\hat{J}_{y}(0)\sin\Omega't.
\end{equation}
If we suppose that initially both wells are equally populated
$\hat{J}_{x}(0)=0$ and the last equation reduces to
\begin{equation}\hat{J}_{x}(t)=\sin(\Omega't)\hat{J}_{y}(0). \end{equation}
As we shall see the dynamical regime imposed by Eq. (19) is
optimal for atomic homodyne  detection.

 By hypothesis we suppose that one of the two modes  of the
condensate (let us say mode B) is prepared in a coherent state
\cite{castin,marcos,nota1} in such way that
$\beta=|\beta|e^{i\theta}$. $\langle\hat{J}_{y}\rangle$ can be
rewritten as
\begin{equation}
\langle\hat{J}_{y}\rangle=\frac{i}{2}|\beta|\left(\langle\hat{a}^{\dag}\rangle
e^{i\theta}-\langle\hat{a}\rangle
e^{-i\theta}\right)=-|\beta|\langle\hat{X}_{\theta-\pi/2}\rangle,
\end{equation} where $\hat{X}_{\theta-\pi/2}$ is the quadrature
operator of the mode A. It is directly seen that
\begin{equation}
\langle\hat{J}_{x}(t)\rangle=|\beta|\sin(\Omega't)\langle\hat{X}_{\theta-\pi/2}\rangle,
\end{equation}
which is the well-known result for balanced homodyne detection
\cite{1a,milburnli} times a coherent amplitude dependent through
$\Omega'$ on the geometry of the trap, the total number of
particles and the initial condition of $\hat{J}_{z}$. It is also
interesting to write the normalized operator $\hat{S}_{i}\equiv\
\hat{J}_{i}/N$ with $i\in (x,y,z)$. The result in Eq. (21) means
that even for large number of atoms, the self-trapping is totally
suppressed and coherent oscillation takes place. We note that in
this regime the frequency of oscillation increases with the total
number of bosons in the system as
$\Omega'=\Omega+2\Lambda(N-1)+8N\kappa\hat{S}_{z}(0)$, in such a
way that the correspondent period decreases. This Rabi regime
allows the double well trap to be envisaged as a realization of a
temporal atomic beam splitter. Hence, the sine function modulating
the homodyne current is an analogue to the beam-splitter
transmissivity factor. For an ideal 50:50 beam splitter the
optimal situation would be such that
\begin{equation}
\Omega' t=(2n+1)\pi/2,
\end{equation}
where $n\in N$. It is preferable to write Eq.(21) as
\begin{equation}
\langle\hat{X}_{\theta-\pi/2}\rangle=\frac{1}{|\beta|}\langle\hat{J}_{x}\rangle
\end{equation}
since the experimentally measured quantity would be the population
difference given by the right-hand side of the above equation.
Remark that since the above derivation is for matter field instead
of the BOHD and thus the quadratures are indeed given by
combinations of the center of mass position and momentum operators
for the mode A relatively to the mode B center of mass. Thus the
atomic homodyne detection would essentially correspond to BEC mode
A center of mass position and momentum measurements.

\section{Atomic Homodyne Detection}
For the complete implementation of the homodyne atomic detection a
measuring process sensitive to the difference of atoms in the two
modes is needed. Here we propose one possible implementation by
letting one of the condensate modes to interact with a far off
resonance cavity light field. The cavity output field is
recombined at a 50:50 beam splitter in a second stage BOHD as
depicted in figure 1. The scheme is very similar to that proposed
in Ref. \cite{7}. One of the wells of the double-well system is
placed in one arm of an optical cavity. The cavity is driven by a
coherent field at the cavity frequency. It is supposed a
dispersive interaction between the light field and the atomic gas
in such a way that the field is far off resonance with respect to
any dipole transition of the atomic species. Hence, the effect of
the atoms is to shift the phase of the cavity field by a
determined amount dependent on the balance of bosons in both
wells. If the atom number in the cavity oscillates, so will the
phase shift. Any tunneling of the condensate will be manifested in
a modulated phase shift of the optical field exiting the cavity.
In order to detect the light phase shift, it is considered a
common BOHD scheme. The light leaving the cavity is thus
recombined with the reference beam in a 50:50 beam-splitter and
allowed to fall on the photodetectors.
\begin{figure}[h] \vspace{-0.3cm}
\includegraphics[width=10cm]{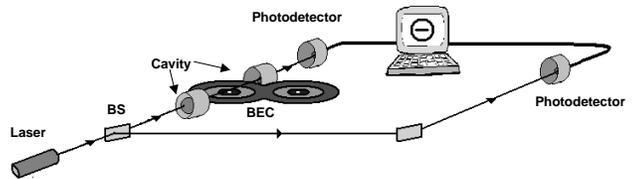}\vspace*{-9cm}
\caption{Atomic homodyne detection scheme. One of the BEC modes
interacts dispersively with the cavity field, after \cite{7}. The
phase shift suffered by the cavity field is dependent on the
imbalance of atoms in the BEC modes. Thus a secondary BOHD on the
cavity output field allows inference of the condensate relative
phase.}
\end{figure}

Throughout the following calculations we assume a bad-cavity
limit, where any related atomic spontaneous emission rate is much
smaller than the cavity field relaxation rate, being thus
neglected. In that regime the cavity field is undepleted and if
the cavity light field is assumed to be far-detuned from any
atomic resonance,  the interaction Hamiltonian is effectively
given \cite{7} by
\begin{equation}
\hat{H}=\int d^{3}r \hat{\Psi}^{\dag}({\bf r})[\hat{H}_{cm}-\hbar\mu
g({\bf r})\hat{c}^{\dag}\hat{c}]\hat{\Psi}({\bf r})
\end{equation}
where $\hat{c}$ and $\hat{c}^{\dag}$ are the cavity field
operators, $g({\bf r})$ is the intensity mode function and
$\mu=\Omega^{2}_{R}/4\Delta$, with Rabi frequency $\Omega_{R}$,
optical detuning $\Delta$ and $\hat{H}_{cm}$ describing the center
of mass motion. We may then write the above Hamiltonian in a
single-particle formalism introducing the condensate operators
$\hat{a}$ and $\hat{a}^{\dag}$ and averaging over the optical mode
function, resulting in the following interaction Hamiltonian
\begin{center}
$\hat{H}_{I}=-\hbar\xi\hat{c}^{\dag}\hat{c}\hat{a}^{\dag}\hat{a}$
\end{center}
\begin{equation}
=-\hbar\frac{N}{2}\xi\hat{c}^{\dag}\hat{c}-\hbar\xi\hat{c}^{\dag}\hat{c}\hat{J}_{x},
\end{equation}
where $\xi$ is the interaction strength. Since the cavity field is
undepleted the total number of photons inside the cavity is a
constant of motion. However the cavity field phase evolves with
time. The phase time evolution can then be found by considering
the Heisenberg equation for the photon annihilation operator if
the undepleted cavity field is assumed to be in a coherent state.
Thus it is direct that
\begin{equation}
\dot{\phi}=-\xi\left(\frac{N}{2}+\langle\hat{J}_{x}\rangle\right),
\end{equation}
showing the direct dependence of the cavity field phase with the
condensate imbalance operator $\langle\hat{J}_{x}\rangle$. If we
now suppose the BEC is being monitored in a balanced homodyne way
as in Fig 1 then it is a well known result that for balanced
homodyne detection schemes, the difference between both fields
arriving at the photodetectors is proportional to the phase in
such a way that
\begin{equation}
\langle\hat{\mathcal{J}}_{x
f}\rangle=-|d|\langle\hat{\mathcal{X}}_{\phi-\pi/2}\rangle
\end{equation}
where $\langle\hat{\mathcal{J}}_{x f}\rangle$ stands for the
photon counting difference at the photodetectors, $|d|$ is the
eigenvalue of the reference beam annihilation operator, and
$\mathcal{X}$ is the cavity field quadrature operator. In this
last equation the light field phase $\phi$ varies with time
depending on the condensate dynamics following Eq. (26). Hence,
the measured photon difference gives us indirect information about
the internal structure of the condensate since it relates itself
directly to the relative phase of the condensate in both wells of
the trapping potential.  A schematic circuit involving both the
atomic and the optical homodyne detection process is depicted in
Fig. 2. It is clear that $\phi$ is a phase shift conditioned on
the number of atoms in the BEC mode inside the cavity.
\begin{figure}[h]
\includegraphics[width=9cm]{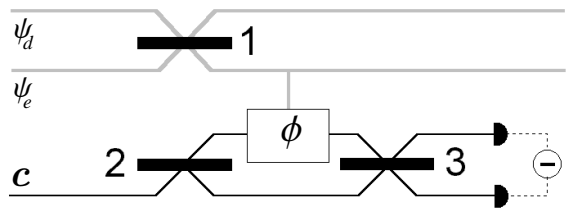}\vspace*{-8.0cm}
\caption{Circuit representing the atomic homodyne detection
process. Grey lines are for the condensate modes and dark lines
for the optical fields. Beam-splitter 1 is the double-well
trapping potential, while  2 and 3 are optical 50:50
beam-splitters. Atoms in the mode inside the cavity interact
dispersively with its field yielding a phase $\phi$ conditioned to
the number of atoms in the mode. The light phase is detected
through BOHD, thus giving information about the relative
condensate phase.}
\end{figure}

We can develop this proposal further by writing
\begin{equation}
\langle\hat{\mathcal{X}}_{\phi-\pi/2}(t)\rangle=
\langle\hat{\mathcal{X}}_{\phi(t)-\pi/2}\rangle=-\frac{i}{2}\left(\langle\hat{c}^{\dag}\rangle
e^{i\phi(t)}-\langle\hat{c}\rangle e^{-i\phi(t)}\right)
\end{equation}
since the field that goes through the other branch of the
beam-splitter does not have \emph{a priori} time dependence, being
$\phi(t)$ as given by Eq. (26). If we assume both light beams
being detected in a coherent state, then
\begin{equation}
\langle\hat{\mathcal{J}}_{xf}\rangle=-|c||d|\sin[\phi(t)].
\end{equation}
We now have an explicit relation between the experimentally
measured quantity ($\langle\hat{\mathcal{J}}_{xf}\rangle$) and the
condensate imbalance operator. In order now to access the relative
condensate phase we need to obtain a relationship between the
quadrature phase operator and the imbalance one. In the following
we find such relationships and provide some real insight on
possible experiments measuring such quantities.

\section{Approximate solutions}

Assuming again the cavity to be driven by a strong coherent field
and being strongly damped, the cavity field is undepleted and in
the $\kappa-eta\ll\Omega'$ limit the Heisenberg equations for the
BEC operators are givenby
\begin{eqnarray}
\dot{\hat{J}_{x}}&=&-\Omega'\hat{J}_{y}-4\kappa[{\hat{J}_{y},\hat{J}_{z}}]_{+},\\
\dot{\hat{J}_{y}}&=&\Omega'\hat{J}_{x}+4\kappa[{\hat{J}_{x},\hat{J}_{z}}]_{+}+\xi\hat{c}^{\dag}\hat{c}\hat{J}_{z},\\
\dot{\hat{J}_{z}}&=&-\xi\hat{c}^{\dag}\hat{c}\hat{J}_{y}.
\end{eqnarray}
In the following we suppose that the BEC in the cavity is strongly
embedded in the photon field in such a way that
\begin{equation}
\epsilon\equiv\frac{\kappa}{\xi N_{f}}\ll1
\end{equation}
where $N_{f}\equiv\langle\hat{c}^{\dag}\hat{c}\rangle$. In other
words we say that the density of the BEC is extremely small
compared to that of the photons in the cavity. It is an
experimental fact that such an approximation is quite correct
since BEC densities range from $10^{12}$ to $10^{13}$
$atoms/cm^{3}$.


In the following, we write down a solution for the set of
differential equations given by Eqs. (21, 22, 23) up to first
order in $\epsilon$ by expanding the Schwinger operators as
$\hat{J}_{i}=\sum_{k}\epsilon^{k}\hat{J}^{(k)}_{i}$. The zeroth
order solution follows directly by integration and it reads
\begin{equation}
\langle\hat{J}^{(0)}_{x}(t)\rangle=\frac{\Omega'}{\omega}|\beta|\langle\hat{X}^{(0)}_{\theta-\pi/2}(t-\pi/2\omega)\rangle,
\end{equation}
where $\omega^{2}\equiv\Omega'^{2}+\xi^{2}N_{f}^{2}$. The first
order solution may be found by considering the homogeneous
solution (zeroth order) and applying the variation of parameters
\begin{eqnarray}
\langle\hat{J}^{(1)}_{x}(t)\rangle&=&\frac{\Omega'|\beta|}{[1+\cos^{2}(2\omega
t)]}\left[\frac{3t}{2}+\frac{1}{4\omega}\cos(2\omega
t)\sin(2\omega t)\right.\nonumber\\
&& \left.-i\frac{3\omega}{{\Omega'}^2}\sin^{2}(\omega
t)\right]\langle\hat{X}^{(1)}_{\theta-\pi/2}(t)\rangle.
\end{eqnarray}
Thus, up to first order in $\epsilon$ the full solution reads
\begin{equation}
\langle\hat{J}_{x}\rangle\simeq\langle\hat{J}^{(0)}_{x}\rangle+\epsilon\langle\hat{J}^{(1)}_{x}\rangle.
\end{equation}
This is a complicated function of time but as it was already
mentioned, it shows that by measuring the imbalance of population
in the wells we acquire information about the relative phase
between both BEC modes as it is expressed in the quadrature phase
operator $\hat{X}_{\theta-\pi/2}(t)$.


We now may plug those results in Eq. (26) and integrate it to
obtain in zeroth order in $\epsilon$
\begin{equation}
\phi^{(0)}(t)=\xi\left(\frac{\Omega'|\beta|}{\omega^{2}}\langle\hat{X}^{(0)}_{\theta-\pi/2}(t)\rangle-\frac{Nt}{2}\right),
\end{equation}
and in first order
\begin{widetext}
\begin{eqnarray}
\phi^{(1)}(t)&=&-\xi\left\{\frac{N}{2}t+\frac{\Omega'|\beta|}{[1+\cos^{2}(2\omega
t)]}\left[\frac{3t^{2}}{4}+\frac{1}{16\omega^{2}}\sin^{2}(2\omega
t)-i\frac{3\omega}{2{\Omega'}^2}\left(t-\frac{1}{\omega}\sin(\omega
t)\cos(\omega
t)\right)\right]\langle\hat{X}^{(1)}_{\theta-\pi/2}(t)\rangle\right\}.
\end{eqnarray}
\begin{figure}[t]
\centering
\includegraphics[width=18cm]{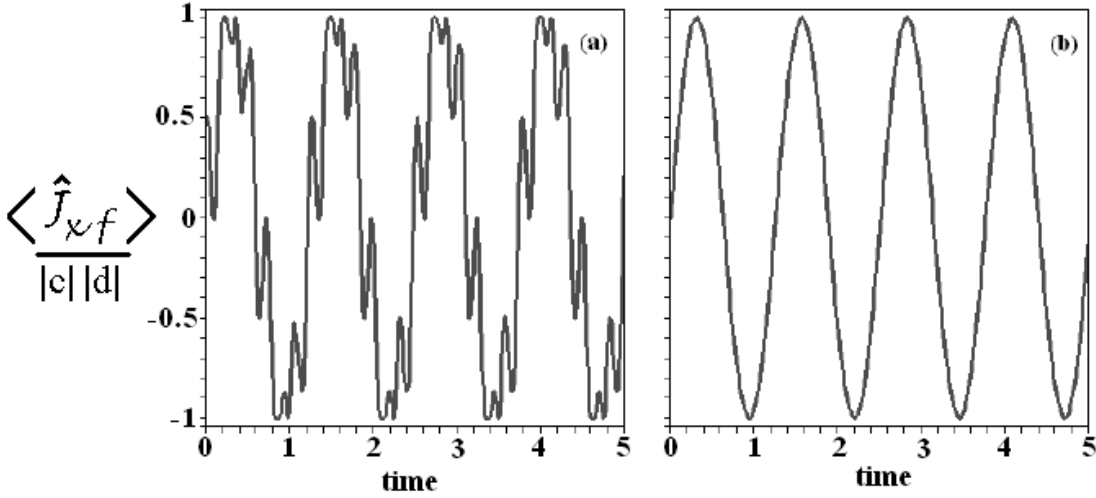}\vspace*{-17cm}
\caption{Time evolution of the homodyne current as shows Eq. (43).
For both graphics $\xi$=0.01,$\Omega'$=25 Hz, $\omega$=30 Hz and
N=10000 atoms. In (a) we suppose a very large initial momentum
(quadrature phase) of $\langle\hat{J}^{(0)}_{y}(0)\rangle$=1667
and in (b) a small initial momentum of
$\langle\hat{J}^{(0)}_{y}(0)\rangle$=0.001.}
\end{figure}
\end{widetext}
Hence, up to first order the full solution reads
\begin{equation}
\phi\simeq\phi^{(0)}+\epsilon\phi^{(1)}.
\end{equation}
As a matter of fact, it is possible to acquire information about
the atomic gas quadrature from the quadrature of the light field
by an atomic temporal homodyne scheme followed by a balanced light
field homodyne detection. Though this solution is quite accurate,
for real experimental data it may suffice to write down the
solution up to zeroth order in $\epsilon$. Consider for instance
$\kappa/\Omega=0.02, \xi=10^{-3} Hz, \Omega=10^{3} Hz,
N_{f}=10^{10}, \Omega'=9 Hz$ and N=1000. For such a situation
$\epsilon\sim10^{-9}$ corroborating a zeroth order approximation
of the problem. Henceforth, the phase $\phi$ reads
\begin{equation}
\phi(t)\simeq\phi^{(0)}(t)=\xi\left(\frac{\Omega'|\beta|}{\omega^{2}}\langle\hat{X}^{(0)}_{\theta-\pi/2}(t)\rangle-\frac{Nt}{2}\right).
\end{equation}
It is interesting however to write down the imbalance in terms of
the light phase since that phase could possibly measured by
experimentalists giving indirect information about the atomic gas
internal structure. In this sense it is easily seen that
\begin{equation}
\langle\hat{X}^{(0)}_{\theta-\pi/2}(t)\rangle\simeq\frac{\omega^{2}}{\Omega'|\beta|}\left(\frac{1}{\xi}\phi+\frac{Nt}{2}\right).
\end{equation}
It is possible now to write down the full zeroth order solution to
the Schwinger operators mean values so to write down an equation
for the light field phase depending explicitly on time:
\begin{equation}
\frac{\phi}{\xi}=-\left(\frac{\Omega'}{\omega^{2}}\cos(\omega
t)\langle\hat{J}^{(0)}_{y}(0)\rangle+\frac{Nt}{2}\right),
\end{equation}
meaning that for sufficiently large initial momentum
$\langle\hat{J}^{(0)}_{y}(0)\rangle$ the harmonic behavior should
dominate and for smaller values the regime should be linear in
time.




 For typical data of the system described here the zeroth order
terms in $\epsilon$ dominate and only coherent oscillations are
observed for the imbalance of the atomic population given by the
mean value of the Schwinger operator $\hat{J}_{x}$. In this
approximation, the relationship between both phases is linear as
we choose specific intervals of time when the potential barrier in
the condensate acts as an ideal 50:50 temporal atomic
beam-splitter. These results make the model here presented an
ideal system for optimal detection of the condensed phase via two
homodyne detections: one temporal atomic one and a second one on
the cavity output light field as depicted in Figs. 1 and 2.

It is now possible to write down an expression which shows us how
the light counting difference at the detectors is directly related
to the internal structure of the condensate. Combining Eqs. (41)
and (42) it is readily seen that
\begin{equation}
\langle\hat{\mathcal{J}}_{xf}\rangle=|c||d|\sin\left[\xi\left(\frac{\Omega'}{\omega^{2}}\cos(\omega
t)\langle\hat{J}^{(0)}_{y}(0)\rangle+\frac{Nt}{2}\right)\right].
\end{equation}
The above expression shows clearly the relationship between the
measured homodyne current and the condensate quadrature phase. In
this sense, it is possible to measure the homodyne current
$\langle\hat{\mathcal{J}}_{xf}\rangle$ and then acquire the
necessary phase info about the condensate via Eq. (42). It is easy
then to see that the evolution of the homodyne current with the
condensate initial momentum (quadrature phase) is simply
sinusoidal and its time evolution is quite similar but with the
envelope function being modulated as shown in Fig. 3 for two
different choices of the initial momentum. The condensate relative
phase is thus more evident for larger initial momenta. Such a
result is evidently optimal for homodyne measurements and reflects
the effective Rabi dynamics discussed previously.

\section{Measurement back-action}
We have considered a measurement process which allows the
inference of a condensate relative phase through optical phase
detection. However as is well known the continuous detection
process induces a back-action into the condensate, altering thus
the condensate phase during the measurement process. We now
analyze those effects by assuming the more realistic situation
 in that the cavity is driven by a strong coherent field of strength
$\varsigma$ and is strongly damped at the rate $\gamma$. The
procedure follows closely that by Corney and Milburn
 \cite{7}. Hence the unconditioned evolution of the system (light
field+ BEC) is governed by the following  master equation (taking
$\hbar=1$),
\begin{eqnarray}
\dot{\hat{\rho}}_{tot}&=&-i[\hat{H}_{I},\hat{\rho}_{tot}]+i\xi[\hat{c}^{\dag}\hat{c}\hat{J}_{x},\hat{\rho}_{tot}]-i(\delta-\frac{N\xi}{2})[\hat{c}^{\dag}\hat{c},\hat{\rho}_{tot}]\nonumber\\
&&-i\epsilon[\hat{c}^{\dag}+\hat{c},\hat{\rho}_{tot}]+\frac{\gamma}{2}\left(2\hat{c}\hat{\rho}_{tot}\hat{c}^{\dag}-\hat{c}^{\dag}\hat{c}\hat{\rho}_{tot}-\hat{\rho}\hat{c}^{\dag}\hat{c}\right),\nonumber\\
\end{eqnarray}
where the initial detuning $\delta=\frac{N\xi}{2}$ was chosen in
order to remove the $N$ linear dependent dispersion.
%
\begin{figure}[h]
\includegraphics[width=9cm]{
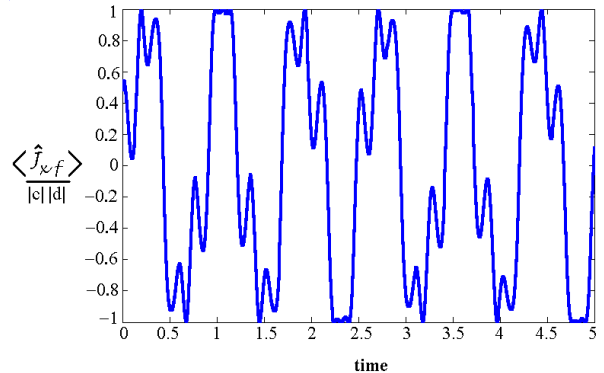}\vspace*{-6.5cm} \caption{Numerical calculation for the
unconditional evolution of the homodyne current when
$\Gamma/\Omega'=0.0065$, $\xi|c_{0}|^{2}/\Omega'=0.04$. Time is
normalized in units of $\Omega'$.}
\end{figure}


 It is possible to eliminate adiabatically the optical field from
the master equation as in Refs. \cite{7,16}, under the hypotheses
that the driving and damping terms dominate over the coupling one.
This process leads to the master equation in terms of the atomic
variables alone
\begin{equation}
\dot{\hat{\rho}}=\frac{-i}{\hbar}[\hat{H_{i}},\hat{\rho}]-\frac{\Gamma}{2}[\hat{J}_{x},[\hat{J}_{x},\hat{\rho}]]+\mathcal{O}(\epsilon_{0}^{3}),
\end{equation}
where the measurement strength is
$\Gamma=16\xi^{2}\varsigma^{2}/\gamma^{2}$ and
$|\frac{\xi|c_{0}|\langle\hat{J}_{x}\rangle}{\gamma}|=\epsilon_{0}\ll1$
(large damping). We may then observe the ensemble-averaged effect
of the measurement in the operator moment equations for
$\eta\sim\kappa$, up to first order in $\epsilon$:
\begin{eqnarray}
\langle\dot{\hat{J}}_{x}\rangle&\simeq&-\Omega'\langle\hat{J}_{y}\rangle,\\
\langle\dot{\hat{J}}_{y}\rangle&\simeq&\Omega'\langle\hat{J}_{x}\rangle+\xi
|c_{0}|^{2}\langle\hat{J}_{z}\rangle-\frac{\Gamma}{2}\langle\hat{J}_{y}\rangle,\\
\langle\dot{\hat{J}_{z}}\rangle&\simeq&-\xi|c_{0}|^{2}\langle\hat{J}_{y}\rangle-\frac{\Gamma}{2}\langle\hat{J}_{z}\rangle.
\end{eqnarray}
Those equations are the same for the case $\kappa=0$ (coherent
oscillation) considered in Ref. \cite{7} but now the oscillation
regime is attained in the presence of self- and cross-collisions,
for $\eta\rightarrow\kappa$ and $\kappa-\eta\ll\Omega'$. We may
now solve numerically this set of equations and with the aid of
Eq.(26) and (29) find numerically the dependence of the homodyne
current with the condensate phase quadrature. In Fig. 4 we depict
the preselected homodyne current as a function of time for
$\Gamma/\Omega'=0.0001$ and $\eta/\Omega'=0.04$. It is observed in
general that even with the light field typical damping of the
preselected state, the phase information of the field (relative to
the condensate quadrature) is still present and shows similar
behavior to that in Fig. (3a). The chosen initial momentum for the
condensate is as such that the oscillatory behavior shown in Fig.4
is due both to the linear term $N\xi t/2$ as well as the term
proportional to the initial momentum in the sine argument of Eq.
(43).
%
\begin{figure}[h]
\includegraphics[width=9cm]{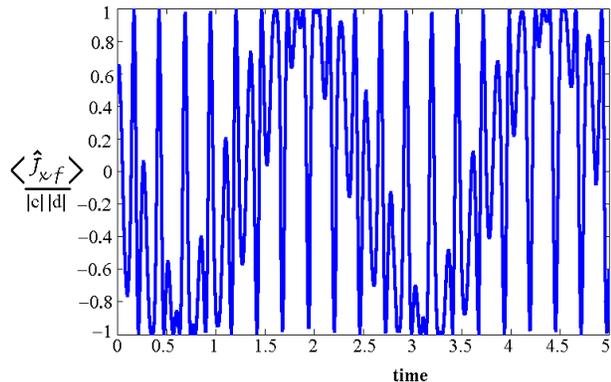}\vspace*{-6.5cm}
\caption{Numerical calculation for the conditional evolution of
the homodyne current when $\Gamma/\Omega'=0.0001$,
$\eta/\Omega'=0.04$. Time is normalized in units of $\Omega'$.}
\end{figure}

We now consider the postselected dynamics of the cavity field plus
condensate system. It is usual then to numerically simulate
stochastic realizations of quantum trajectories as  already
pointed out by several
authors\cite{gardiner,parkins,18,17,gardinerbook}. The resultant
stochastic process is a diffusive evolution rather than the jump
processes, which occur in the direct detection of atoms or
individual photons since we have a condensate system continuously
monitored by the optical homodyne detection scheme. In the
presence of cross-collisions the conditional master equation
(which corresponds to an average over many runs of the experiment
and many homodyne current records) for the optical field is given
by
\begin{equation}
\left(\frac{d\hat{\rho}_{c}}{dt}\right)_{\mbox{\it
field}}=\gamma\mathcal{D}[c]\hat{\rho}_{c}+\sqrt{\gamma}\frac{dW(t)}{dt}\mathcal{H}[c]\hat{\rho}_{c},
\end{equation}
where dW(t) is the infinitesimal Wiener increment
\cite{gardinerbook}, $\hat{\rho}_{c}$ is the density matrix that
is conditioned on a particular realization of the homodyne current
up to time $t$ and $\mathcal{D}$, $\mathcal{H}$ are the Wiseman's
superoperators \cite{17}. It follows then the conditional
stochastic Schor\"odinger equation
\begin{equation}
d|\tilde{\Psi}_{c}(t)\rangle=dt[-i\hat{H}-\frac{\Gamma}{2}\hat{J}_{x}^{2}+I(t)\hat{J}_{x}]|\tilde{\Psi}_{c}(t)\rangle,
\end{equation}
where $\hat{H}$ is given by Eq.(10) plus Eq.(25) and
$\tilde{\Psi}_{c}(t)$ describes the conditional state of the
system. The measured photocurrent is\begin{equation}
I(t)=2\Gamma\langle\hat{J}_{x}\rangle_{c}+\sqrt{\Gamma}\mathcal{A}(t),\end{equation}
where the stochastic term $\mathcal{A}(t)$ has the correlations
$\langle\mathcal{A}(t)\rangle=0$ and
$\langle\mathcal{A}(t),\mathcal{A}(t')\rangle=\delta(t-t')$.

From Eq. (50) we see that the presence of cross-collisions in the
$\eta\rightarrow\kappa$, ($\kappa-\eta\ll\Omega'$) limit
introduces only an harmonic correction due to the
$\eta\hat{J}_{z}^{2}$ term in the Hamiltonian. Then again, the
cross-collisions reproduce the homodyne interference pattern
expected when there are no collisions at all ($\kappa=0$).
Numerical simulations then show similar results to that found by
Corney and Milburn \cite{7} for only coherent oscillation dynamics
($\kappa=0$) corroborating the fact that such system in the
effective Rabi regime is optimal for homodyne detection in the
sense that it attains a purely coherent oscillation of population
dynamics without collapse and revival. The numerical simulations
depicted in Fig 5 shows as well, how the experimental measured
quantity given by Eq. (29) evolves in time under the conditioned
detection for $\Gamma/\Omega'=0.0001$ and $\eta/\Omega'=0.04$. We
then observe that the homodyne current changes considerably in
relation to the oscillatory unconditioned evolution. This is
expected since the measurement back action alters considerably the
whole system state. An efficient detection process with a larger
initial condensate momentum would allow the evidence of the
condensate phase. It may then be possible to experimentally
confirm such results by measuring the homodyne current and then
inverting Eq. (29) in order to access the imbalance of population
between both wells and then  to acquire the desired information
about the relative quadrature phase of the Bose-Einstein
condensates.

\section{Concluding Remarks and General Discussion}

We have shown that in the effective Rabi regime of a double-well
atomic BEC \cite{bruno} an optimal condition for atomic homodyne
detection scheme is found, which gives indirect measurement of the
condensate relative phase. The double-well potential barrier acts
as a temporal atomic beam-splitter with the transmissivity factor
varying with time and depending directly on the total number of
bosons and cross-collisions strength by the corrected frequency
$\Omega'$ \cite{ijmpb}. Up to first order in $\epsilon$, the
Heisenberg equations of motion for the mean values of the
Schwinger operators are exactly soluble even when the interaction
with the light field is considered strong for sufficiently strong
light intensities.

Typical experimental data show that it may suffice to consider
only zeroth order terms in the calculations which result in a
linear relationship between the light phase and the condensate
quadrature. In this sense it is supposed that the light phase may
be detected  with the aid of the scheme proposed in Figs. 1 and 2.
It consists of a two stage homodyne detection, one optical and the
other on the state of one of the two-mode condensate. Hence, we
believe that such a system in this dynamical regime (effective
Rabi) might be the appropriate choice to indirectly detect the
relative phase between the two modes of a BEC in a double well
potential in the form of Josephson-like tunneling in a regime of
purely coherent exchange of population between both wells due to
the strong presence of cross-collisions. Such a conclusion is
strongly supported by the calculations (analytical and numerical)
discussed in this paper.

Recently an outstanding experiment was realized based on
stimulated light scattering to continuously sample the relative
phase of two spatially separated atomic BECs that never interact
\cite{saba}. Our proposal on the other hand imposes that the two
atomic BEC modes must be overlapping in order that the effective
(stable) Rabi regime ($\kappa-\eta\ll\Omega'$) be attained. In
face of Eq. (26) and (29), we expect experimentalists to be able
to measure the relative phase with present technologies on
trapping potentials. Experiments should then be able to detect the
relative condensate phase, as given by the model here presented,
possibly opening new frontiers in quantum phase engineering. We
expect that these results may be useful in further experimental
and theoretical studies on the state of a Bose-Einstein condensate
as well as to applications on atom optics when such systems are
extrapolated to an array of BECs. As a last comment, our results
are certainly relevant for
 reconstruction and measurement of atomic quantum
states \cite{8,9,10} and may be useful for future implementations
on quantum communication protocols \cite{marcos,qtbox}.

\begin{acknowledgements}

The authors would like to acknowledge partial financial support
from FAPESP under project $\#04/14605-2$, from CNPq and from
FAEPEX-UNICAMP.
\end{acknowledgements}


\begin{thebibliography}{0}
\bibitem{nota1}
 The quantum state of Bose gases is a
central problem and shall not be addressed here. We simply assume
that one of the condensates can be found close to a coherent state
to simplify our calculations. For a good discussion on that see
Refs. \cite{leggett,walls}.
\end{thebibliography}
\end{document}